Patterns of Research Funding Across Research Subjects: The Case of NSERC

Rabeeh Parhizkari[*] and Gita Ghiasi[**]
[*]Rparh078@uottawa.ca; gghiasi@uotatwa.ca
0009-0003-1602-7232; 0000-0001-8296-6674

School of Engineering Design and Teaching Innovation, University of Ottawa, Canada

**Abstract**
Funding agencies aim to maximize scientific output from the resources they allocate, a goal that may inadvertently reinforce cumulative advantage within funding systems. Focusing on research areas within Natural Sciences and Engineering (NSE), this study examines the distribution of Canadian research funding over time, with particular attention to the extent to which certain subjects gain prominence as emerging or high-growth fields, while others experience relative stagnation or decline. The results indicate that many research subject groups receiving decreased funding in recent years are concentrated in the pure sciences and mathematics. In contrast, social sciences and medical sciences, which initially had relatively low average funding per grant application, have demonstrated increasing funding trends over time. These findings offer valuable insights for funding agencies seeking to design policies that promote more effective and equitable allocation of resources.

**Introduction**
Funding grants have increased globally over the past decades, as documented in countries such as Denmark, Norway, the USA, Canada, and the European Commission (Duan, 2022; Huang & Xiong, 2025; Huang et al., 2016; Jin et al., 2026; Wu, 2015). Also, the average grant size in China has increased from 0.18 million yuan in 2001 to 0.72 million yuan in 2020 (Duan, 2022). Even though there is an increasing trend in grant funding across most countries, funding patterns vary significantly across regions, suggesting that these variations may depend on national priorities rather than on global trends alone. For example, research funding from the Taiwan Government from 1999 to 2009 was almost stable (Huang, 2021). This contrast illustrates how national contexts play a role in shaping funding trajectories differently across countries.
In addition to international variation**,** research funding has not only experienced an upward trend in universality (increasing from 66.30% to 74.26%) but also in multiplicity (increasing from 2.82 to 3.26) (Tian et al., 2024). Moreover, the proportion of super-funded articles (articles with more than three projects) grew from 24.93% in 2011 to 32% by 2020 (Tian et al., 2024). These findings indicate broader structural changes in funding practices.
Turning to the issue of equity, studies have also examined how evenly research funding is distributed across researchers, institutions, and disciplines. For example, Wu (2015) showed a general decrease in funding inequality across 1,971 universities from 2000 to 2013, as measured by both funding volume and grant counts. However, despite reductions in inequality in some contexts, research funding has become more competitive and performance-based in recent years, with agencies increasingly interested in evaluating funded projects after completion (Mutz et al., 2012). Studies show that institutions or individuals who have received funding before are more likely to get funding (Huang, 2021; Mongeon et al., 2016), and researchers who begin at higher funding levels are more likely to maintain their position over time (Wahls, 2018). This might be because funding agencies aim to maximize scientific output from the funds they provide, which collectively reinforces the persistence of cumulative advantage within funding systems.
Consequently, the literature has increasingly raised concerns about funding inequality. Empirical evidence suggests that funding inequality has been rising since 1985 (Wahls, 2018), with a small proportion of researchers and institutions receiving a disproportionately large share of available funding (Wang & Wu, 2023). Publications supported by funding outperform unfunded publications across measures of variety and disparity (Huang & Xiong, 2025). Looking at the scientific output, evidence suggests that research productivity and impact do not increase linearly with higher levels of funding; instead, beyond a certain threshold, additional resources are associated with diminishing marginal returns in both productivity and impact (Wahls, 2018). Similar patterns have been observed in the Canadian context, where Mongeon et al. (2016) found that concentrating funding among a small group of elite researchers results in diminishing returns, with highly funded researchers not significantly outperforming their peers in terms of output or scientific impact. Furthermore, Wahls (2018) demonstrated that less-prestigious institutions produced 65% more publications and achieved 35% higher citation impact per dollar of funding. Collectively, these findings suggest that distributing smaller grants to a broader pool of researchers may enhance overall research productivity while addressing concerns about funding inequality.

With particular attention to research subjects in Natural Sciences and Engineering (NSE), this study examines the concentration of Canadian funding over time, focusing on the extent to which certain subjects gain prominence (emerging or high-growth areas) while others experience relative stagnation or decline, to help funding agencies develop policies that promote more effective and equitable resource allocation. Understanding these allocation patterns is essential for policymakers, researchers, and society at large. We further extended our focus to different types of NSERC grants: namely, the Discovery Grants Program, the Partnerships Program, and the Training Programs. Discovery Research is awarded to advance foundational knowledge and to long-term, curiosity-driven research programs rather than to short, predefined projects. Research Partnerships Programs are awarded to support collaborations among academia, industry, government, and non-profit organizations. Training Programs focus on training Highly Qualified Personnel, including undergraduate, graduate, and postdoctoral researchers.

## Method

Initially, we ensured that the names and codes of subject areas were consistent across years. To do this, we grouped all areas with the same name but different codes, and all codes that were identical but had different names and assigned a single name and code to each. Thus, in the new dataset, each code was assigned to exactly one name, and each research subject group was assigned a unique code. The funding types are divided into three categories: the Discovery Grants Program, the Partnerships Program, and the Training Programs. The purpose of each of these awards is different. Our analysis is filtered by the types of awards granted to researchers. According to NSERC, Discovery Research is awarded to advance foundational knowledge and to long-term, curiosity-driven research programs rather than to short, predefined projects. Research Partnerships Programs are awarded to support collaborations among academia, industry, government, and non-profit organizations. Training Programs focus on training Highly Qualified Personnel, including undergraduate, graduate, and postdoctoral researchers.

Since the purposes of these awards are very different from one another, we filtered the funding data by grant group and calculated the trends separately. Because the awards granted in the Training Program, Research Partnerships, and Discovery Programs have completely different objectives, separating them allowed us to perform a more accurate analysis.

We calculated trend score parameters to examine which research areas have trended over the past 24 years (2001-2024) and which have declined. To calculate the trend score of research subjects, we use a logarithmic growth rate commonly used in studies for calculating the changes of distribution over time. This approach gives us a more accurate representation of the long-term trend. The parameter is designed to handle the number of projects, the total funding amount, and the early- and late-funding amounts allocated to that field. By doing so, we aimed to design a parameter that is robust to the number of years a research field has existed. We incorporate inactive years to make our trend score robust to periods without funding. There is only one research subject group in the dataset that is followed by a gap before funding resumes after 2008, and that is namely Social Sciences and Humanities. To ensure robustness and focus on how the support changed over the active years, only these years are included; including inactive years would substantially bias the Trend Score for those years and would obscure the underlying pattern. Ultimately, the parameter we developed shows how the funding for a research subject has changed from the first year it received funding to the most recent years.

Funding Trend Parameter: For each subject area $s$ and year $t \in \{2001, \ldots, 2023\}$, let

– $F_{s,t}$ denotes the total funding amount allocated to subject s in year t,
– $P_{s,t}$ denotes the total number of funded projects for subject s in year t.
Let $T_s \subseteq \{2001, \ldots, 2023\}$ denote the set of years in which subject s is present:
$T_s = \{ t \mid F_{s,t} > 0 \text{ or } P_{s,t} > 0 \}$.
Define the first and last observed years of the subject s as:
$t_{min}(s) = \min T_s$ , $t_{max}(s) = \max T_s$ .
To quantify how funding and project activity change over the entire observed period, we compare the averages in the early timeline of the observed period and the late timeline. Let
$m_s = |T_s|$ and define:

$$T_s^{early} = \{\text{the first } (\max(1, \lfloor \frac{m_s}{2} \rfloor)) \text{ years in Ts}\},$$
$$T_s^{late} = \{\text{the last } (\max(1, \lfloor \frac{m_s}{2} \rfloor)) \text{ years in Ts}\}.$$

For each subject s, define the corresponding averages, where $|T_s^{early}|$, $|T_s^{Late}|$ denote the number of years in the early and late timeline.

$$\bar{F}_{s,E} = \frac{1}{|T_s^{early}|} \sum_{t \epsilon T_s^{early}} F_{s,t} \qquad \bar{F}_{s,L} = \frac{1}{|T_s^{Late}|} \sum_{t \epsilon T_s^{Late}} F_{s,t}$$

$$\bar{P}_{s,E} = \frac{1}{|T_s^{early}|} \sum_{t \epsilon T_s^{early}} P_{s,t} \qquad \bar{P}_{s,L} = \frac{1}{|T_s^{Late}|} \sum_{t \epsilon T_s^{Late}} P_{s,t}$$

Finally, the funding trend score would be calculated:

$$T(s) = \alpha \log(\frac{\bar{F}_{s,L} + \varepsilon_F}{\bar{F}_{sL}E + \varepsilon_F}) + (1-\alpha) \log(\frac{\bar{P}_{s,L} + \varepsilon_P}{\bar{P}_{s,E} + \varepsilon_P})$$

**Results**

Figure 1 shows that total funding has increased over the years; however, the number of projects has not been stable over the years. In this research, the term project means a granted application. Even though studies suggest that increasing the number of funding allocations would have better and more efficient results than concentrating the funding on a small group of researchers (Mongeon et al., 2016).

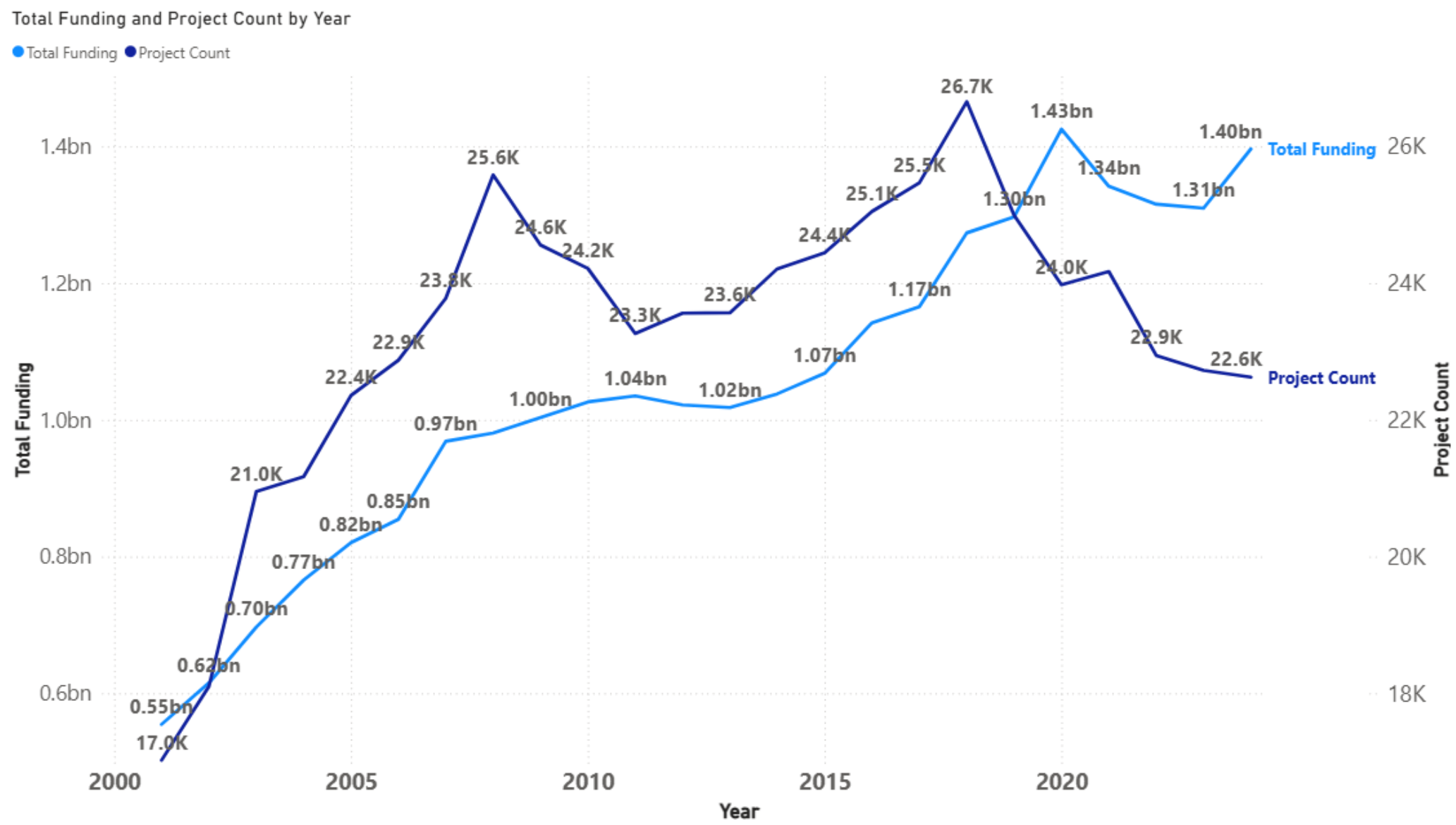

Figure 1: Number of grants growth vs average funding over the last 24 years

Figure 2 presents a heatmap of the average project funding per subject group for NSERC research over the past 24 years.

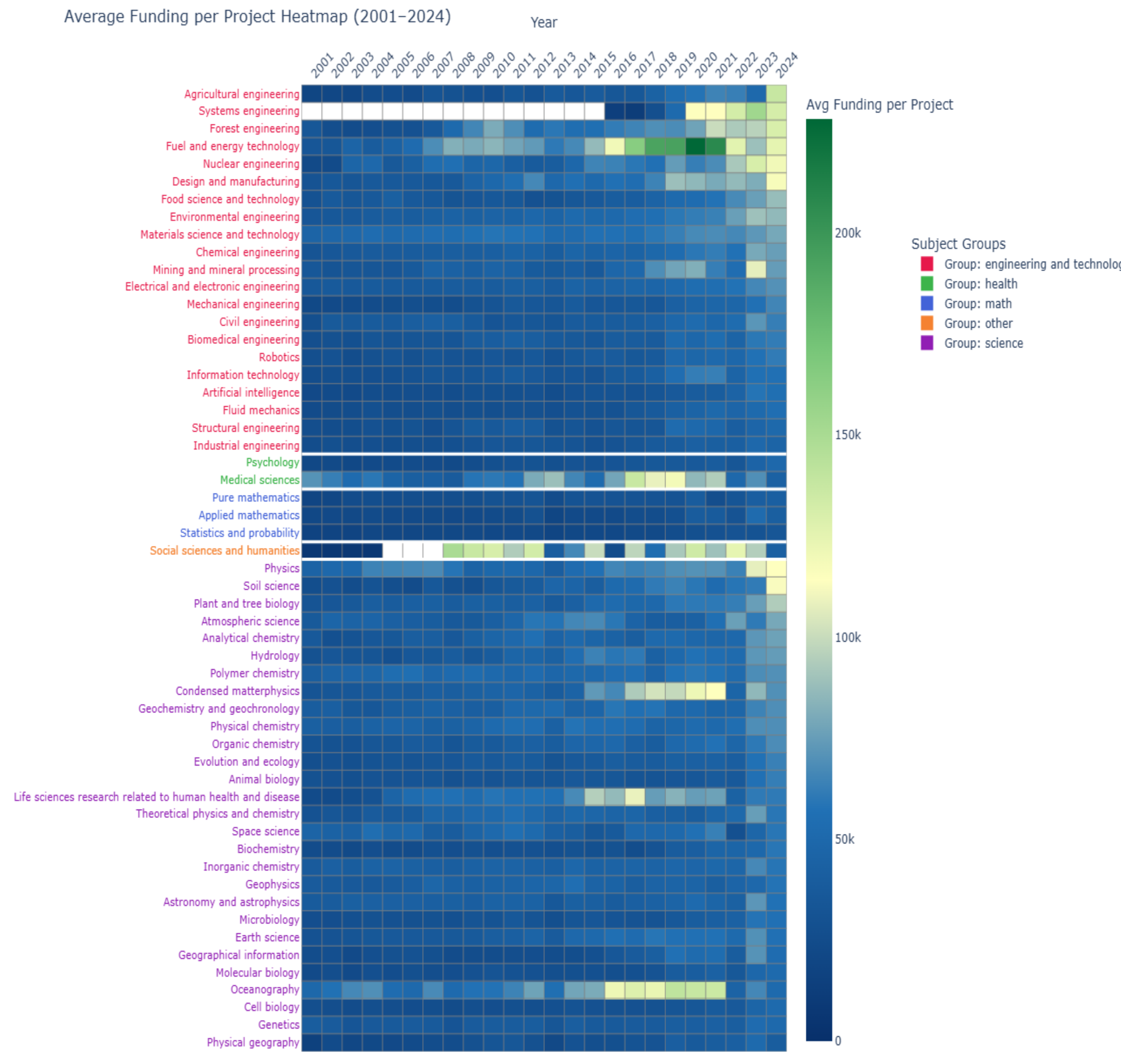

Figure 2: Heatmap for average funding per project for all subject groups

This heatmap reveals several notable and uneven patterns: Some subject areas show a gradual increase in average funding over time, including systems engineering, forest engineering, fuel and energy technology, nuclear engineering, and design and manufacturing. In contrast, other fields exhibit more variable funding trends. Medical sciences, condensed matter physics, life sciences, and oceanography show fluctuating patterns, with relatively low funding in the early years, followed by periods of higher funding and subsequent declines in more recent years. Oceanography, in particular, displays a distinct increase in average funding from 2016 to 2021 and appears to have received exceptionally less funding in 2023 and 2024. Fuel and energy technology received higher average funding from 2017 to 2021 but experienced a decline in 2022 and 2023. Finally, systems engineering has received progressively higher average funding from 2021 through 2024. The social sciences exhibit an exceptionally fluctuating funding pattern, likely driven by a small number of large funding awards in certain years. However, this observation requires further detailed investigation.

Figure 3 shows the five research subject groups with the highest trend scores and the five with the lowest trend scores. Among the research subject groups with the highest trend scores and high average funding are Fuel and Energy Technology, Medical Sciences, and Social Sciences and Humanities. In contrast, Organic Chemistry, Theoretical Physics and Chemistry are among the research subject groups with low average funding and the most negative trend scores.

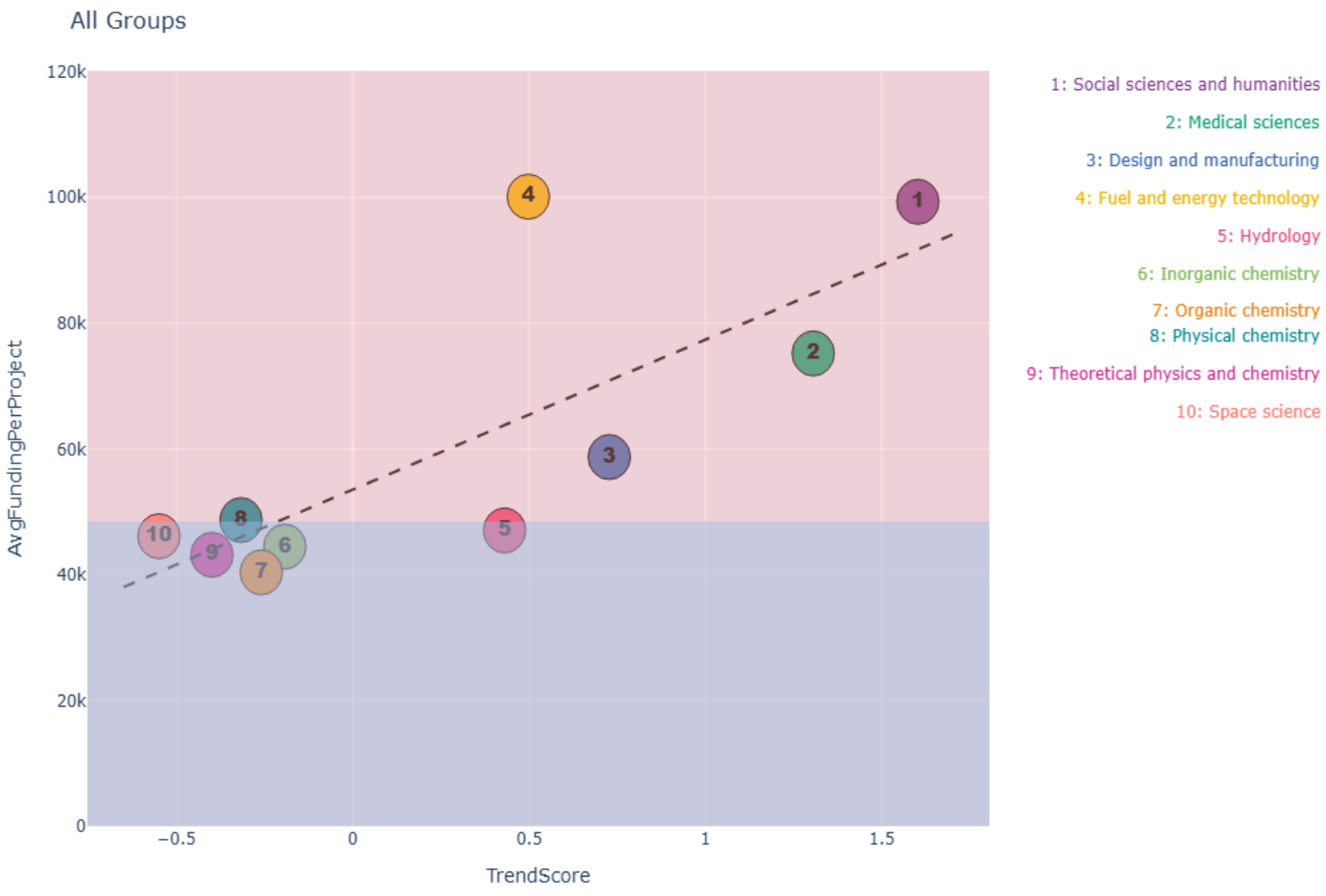

Figure 3 – Trend Score vs. Average NSERC funding per year for funding across subject groups

Tables 1 and 2 show the Trend Scores by grant type (Discovery, Partnership, and Training) and for overall NSERC funding among the five research subject groups with the highest Trend Scores and the five with the most negative Trend Scores. Medical Sciences is among the most highly trending research subject groups across all grant types and in overall NSERC funding. In contrast, Space Sciences is among the research subject groups with the most negative Trend Scores across all grant types and in overall NSERC funding. Another research subject group of particular interest is Fuel and Energy Technology. Although it ranks among the research subject groups with the highest overall NSERC funding Trend Scores, it has the negative Trend Score for the Training program while exhibiting positive Trend Scores for every other program. This may indicate a decline in funding for training students in this field despite increasing funding through the Discovery and Partnership programs. Furthermore, among the research subject groups with the most negative Trend Scores, four had positive Trend Scores in the Partnership program: Inorganic Chemistry, Organic Chemistry, Physical Chemistry, and Theoretical Physics and Chemistry. This may indicate increasing industry partnerships in these fields despite more limited funding through the Training and Discovery programs. Overall, pure science research subject groups predominantly exhibit negative Trend Scores, whereas engineering research subject groups predominantly exhibit positive Trend Scores.

Table1: Trend Scores for Research Subject Groups, with the highest Trend Score in all programs combined

| Research subject group | All | Training | Partnership | Discovery |
|---|---|---|---|---|
| Social Sciences and Humanities | 1.60 | 0.04 | 1.74 | 0.23 |
| Medical sciences | 1.30 | 1.24 | 1.50 | 1.17 |
| Design and manufacturing | 0.72 | 0.44 | 1.35 | -0.09 |
| Fuel and energy technology | 0.49 | -0.45 | 0.55 | 0.92 |
| Hydrology | 0.43 | -0.07 | 0.64 | 0.53 |

Table2: Trend Scores for Research Subject Groups with the lowest Trend Score in all programs combined

| Research subject group | All | Training | Partnership | Discovery |
|---|---|---|---|---|
| Inorganic chemistry | -0.19 | -0.24 | 0.20 | -0.15 |
| Organic chemistry | -0.25 | -0.54 | 0.33 | -0.16 |
| Physical chemistry | -0.31 | -0.51 | 0.19 | -0.009 |
| Theoretical physics and chemistry | -0.39 | -0.88 | 0.81 | -0.01 |
| Space science | -0.54 | 0.04 | -0.18 | -0.03 |

**Discussion and conclusion**

This study aimed to understand longitudinal changes in funding over the past two decades. Specifically, we evaluated whether funding allocated across different NSERC research subject groups is distributed unevenly.

The findings indicate that although total funding has increased over the years, the number of projects has declined, especially after 2017. This may indicate that funding is becoming increasingly concentrated, with resources allocated to a smaller group of researchers. In addition, the analysis indicates that more pure science research subject groups exhibit a negative Trend Score than applied science and engineering research subject groups. This suggests a shift in research focus from the pure sciences toward applied science and engineering research subject groups.

We also introduced a Trend Score to guide future studies investigating how a parameter changes over time and whether it has been increasing or decreasing in recent years. To our knowledge, this is the first application of such a score in this context. This metric is designed to be robust to both the number of years and the magnitude of the parameter.

We identified the research subject groups with the strongest upward funding trends and the most pronounced downward funding trends for each grant program. Medical Sciences exhibited a high Trend Score across all programs, whereas Space Sciences exhibited the lowest combined Trend Score across all programs. The declining funding trend for Space Sciences may need policy attention. The analysis further highlights research subject groups that have a positive Trend Score in one or two programs and a negative Trend Score in others. Specifically, for research subject groups such as (In)Organic Chemistry, Physical Chemistry, Theoritical Physics and Chemistry exhibit an upward trend in Partnership funding but a downward trend in Discovery and training funding, it may reflect increasing industry relevance.

The Trend Score uncovered several patterns in the distribution of funding over the past 24 years. For instance, we found that some research subject groups have received less funding in recent years. These declining funding levels could be considered in future policymaking. In future work, we plan to extend this study using datasets from the Canadian Institutes of Health Research (CIHR) and the Social Sciences and Humanities Research Council (SSHRC) to further our goal of identifying research fields with declining funding trends.

## Author contribution

**Rabeeh Parhizkari:** Methodology, Visualization, Investigation, Data curation, Writing- Original draft preparation, **Gita Ghiasi.**: Conceptualization, Investigation, Supervision, Reviewing and Editing.